\UseRawInputEncoding
\documentclass[12p,sort&compress]{elsarticle}

\usepackage{lineno,hyperref}
\usepackage{graphicx,epsfig,amscd,amsmath,amssymb,float}
\usepackage{natbib,setspace,pbox,makecell}
\journal{Carbon}

\begin{document}

\begin{frontmatter}
\title{Doping of Large-Pore Crown Graphene Nanomesh} 
\author[zewail]{Mohamed S. Eldeeb}
\author[zewail,benha]{Mohamed M. Fadlallah}
\author[ibm]{Glenn J. Martyna}
\author[iau]{Ahmed A. Maarouf\corref{cor1}}
\ead{amaarouf@iau.edu.sa}

\cortext[cor1]{Corresponding author}

\address[iau]{
Department of Physics, Institute for Research and Medical Consultations, Imam Abdulrahman Bin Faisal University, Dammam 31441, Saudi Arabia}

\address[zewail]{Center for Fundamental Physics, Zewail City of Science and Technology, Giza 12588, Egypt}

\address[benha]{Department of Physics, Faculty of Science, Benha University, Benha, Egypt}

\address[ibm]{IBM T. J. Watson Research Center, Yorktown Heights, NY 10598, USA}

\begin{abstract}
Graphene nanomeshes (GNM's) are garnering increasing interest due to their potential application to important technologies such as photovoltaics, chemical sensing, ion-filtration, and nanoelectronics. Semiconducting GNM's with fractional eV band gaps are good candidates for graphene-based electronics, provided that a mechanism for their stable and controlled doping is developed. Recent work has shown that controlled passivation of the edges of subnanometer pores and subsequent doping give rise to {\it n}- and  {\it p}-doped GNM structures. However, these structures are difficult to fabricate at the nanoscale. Here, we use first principle calculations to study the effect of the pore size on the doping physics of GNM structures with larger pores that can potentially host more than a single dopant. We show that such doping mechanism is effective even for pores with relatively large radii. We also study the effect of the number of dopants per pore on doping stability. We find that stable rigid band {\it n}- and {\it p}-doping emerges in such structures even if the dopants form a nano-cluster in the pore - rigid band doping is achieved in all cases studied. 
\end{abstract}

\end{frontmatter}


\section{Introduction}

The fascinating physical and chemical properties of graphene make it a unique material that has attracted a lot of attention over the last decade. Most recently, its derivative materials are being considered for potential use in many technological applications, such as nanoelectronics,\cite{Yang2017,Withers2015} molecular separation,\cite{aza2017,Huang2015} and catalysis. \cite{Hu2017,Fan2015}
 
A single-layer graphene sheet has a transparency of about $98\%$.\cite{graphenetransparency} This nominates graphene for use as a transparent electrode material.\cite{Mu2016,Maaroufhybrid2016} One problem is the relatively high sheet resistance of pristine graphene, which necessitates its doping, by moving the Fermi level away from the Dirac point. Graphene doping is, in general, unstable, primarily because of the relative chemical inertness of graphene arising from its {\it sp}$^2$ bonding. Dopants tend to physisorb on its surface, rendering the lowering of the resistance only temporary. Without the development of a stable doping method for graphene, its utilization as a transparent electrode will be greatly limited. 

Despite its high electronic mobility (upon doping),\cite{Bolotin2008} graphene lacks an electronic gap, which limits its potential as a substitute for silicon in the transistor world. Few attempts to open a gap in single and multilayer graphene have been made, for example, by exploiting quantum confinement by forming graphene nanoribbons,\cite{nanoribbons}, by subjecting a bilayer to an orthogonal electric field.\cite{electricfield}, or by subjecting graphene to strain.\cite{strain2} All routes come with some challenges, and thus are not technologically feasible, hindering the fabrication of a graphene-based field effect transistor (FET).

Graphene nanomeshes (GNM's, also called porous graphene) may come to the rescue by solving both of graphene's problems by providing the means (1) to overcome the relative chemical inertness, hence allowing for doping stability, and (2) to open a technologically appealing bandgap, providing graphene with its pass to the transistor world. GNM's are crown-like structures formed by creating a super-lattice of pores in a graphene sheet. The pore edges boost the graphene chemical activity, permitting it to bind with appropriate chemical dopants. Furthermore, and depending on the super-lattice constant and the pore geometry, these structures can inherit graphene's semimetallicity, or can be semiconducting with a fractional eV gap.\cite{Martinazzo2010} If such semiconducting structures could be both controllably and stably doped, they could be used to fabricate graphene-based transparent electrodes, computer logic switches, and spintronic devices.\cite{Fadlallah2016}

\begin{figure}[H] 
\begin{center} 
   \includegraphics[width=2.5in]{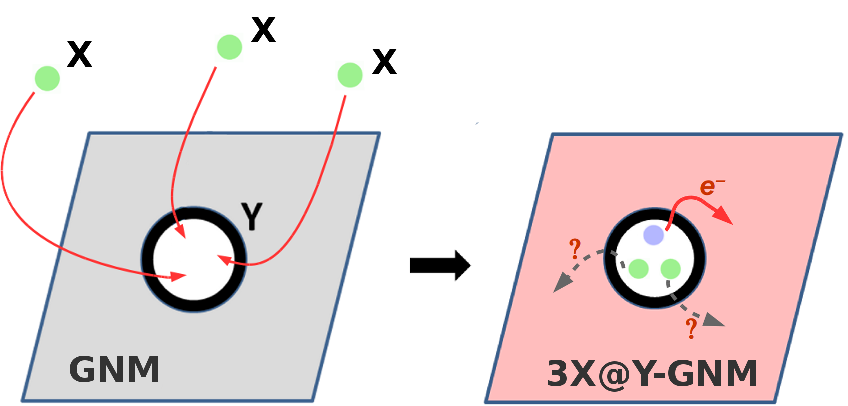}
\caption{A conceptual sketch of multiple-ion chelation doping of a GNM. Chelant atoms are successively brought close to pore, where some or all of them become ionized, thus doping the GNM. Here, {\it n}-doping is shown; {\it p}-doping simply replaces the electron with a hole. In principle, partial charge transfer can occur, where the HOMO/LUMO of the chelants can move relative to the GNM spectrum.}
\label{model}
\end{center} 
\end{figure}

A method for the controlled, stable, chemical doping of pore-edge passivated GNM's has been recently proposed. \cite{Maarouf2013,maaroufdopingpatent} A neutral dopant undergoes a charge transfer reaction with the GNM of specific pore edge passivation, ionizes, and is then electrostatically trapped in the pore by the local dipole moments of the edge functional groups. The charge so transferred dopes the GNM, moving the Fermi level into the valence or the conduction bands (see Fig. \ref{model}). The GNM band structure is essentially not affected by the doping process, and the dopant state involved in the charge transfer is far from the Fermi level. In this way, a stable controlled rigid band doping of semiconducting GNM's can be achieved, at least for those designed to accommodate a single dopant.\cite{Maarouf2013,maaroufdopingpatent}

GNM's have been fabricated by several groups, with pore-size distributions in the 3-200 nm range.\cite{konig2016self,C4NR04584J} A study of the I-V characteristics of {\it p}-doped GNM-based transistors indicated that their ON-OFF ratio is an order of magnitude larger than that of pristine graphene, but with lower electrical conductivity. Most recently, sub-nanometer pores were fabricated, albeit without the periodicity necessary for transport application.\cite{Guo2014} GNM's have also been considered for various applications, including supercapacitors and metal-free electrodes. \cite{C4NR04584J, 6b09836, C6TA09734K, C6NR01838F} Since fabrication techniques are likely to yield GNM's with nanometer-sized pores which can host more than one dopant atom, some questions arise regarding the dopant stability in the pores. How would multiple dopants behave in a large pore? Would they uniformly distribute around the pore edge, or would they cluster together, with obvious impact on the dopants binding energy? In addition, how would the distribution/clustering of the dopants in the pore affect the rigid band doping picture?

In this work, we address the aforementioned questions through the study of the electronic properties of a representative set of GNM structures with varying pore sizes, dopant loads, and pore edge passivations, for both {\it n-} and {\it p}-doping. Using first principles calculations, we calculate the binding energy as well as the maximum number of dopants bound per pore. We also explore the effect of multiple small dopants on the band structure of the doped GNM's to determine whether or not the rigid band picture is applicable, and if there is a multi-ion doping state that will emerge at or close to the Fermi energy. The latter point is key for transistor applications, but irrelevant applications that only require high mobility and carrier density. Multi-ion doping can be avoided by designed complex molecules that fit into the large pores - a topic to be described in future work. 

\section{Computational Methods}
Electronic structure calculations were performed with the Quantum Espresso software package \textregistered. Minimum energy configurations were obtained by relaxing the X@Y--GNM$c$ systems with forces less than 0.001 Ry/Bohr. The calculations were performed using the generalized gradient approximation PBE functional, \cite{PBE1} with an energy cutoff of $45$ Ry, a $12 \times 12 \times 1$ Monkhorst-Pack k-point grid for the X@Y-GNM9 systems, and a  $6 \times 6 \times 1$ k-point grid for the larger X@Y-GNM12 ones. Our choice of the k-point grid and energy cutoff exceeds commonly used ones in similar systems. \cite{Sun2017,Levita2016} Van der Waals interactions (vdW) were calculated within the semi-empirical DFT-D2 \cite{Dion2004} and the first-principles vdW-DF methods.\cite{Thonhauser2007,grimme,Thonhauser2015} A vacuum distance of $12$ \AA\ was used for the image separation in the direction perpendicular to the plane of the GNM. Charge transfer was calculated by integrating the total density of states.
 
\section{Description of the Physical System}

Since {\it n}- and {\it p}-type FET's are required for digital electronics, the adoption of GNM's for post-CMOS technologies relies heavily on a thorough understanding of their properties, including doping. Experimental work on GNM's currently achieves pore diameters as small as $3$ nm.\cite{zhang2016facile} Therefore, we have to consider multiple dopants per pore, studying the details of their packing and stability for near term applications. These studies also provide a framework for understanding the behavior of dopant ionic complexes for chemical separation applications.

 We have recently proposed to dope GNM's \cite{Maarouf2013,maaroufdopingpatent} based on a concept from chemistry - ion chelation. Here, the pore perimeter is passivated by a species with an electronegativity different from carbon, which results in the formation of local radial dipoles at the pore perimeter creating a favorable environment to host ions. A neutral atom is ionized by the sheet through a charge transfer reaction in which the electrostatic energy gain is higher than the charge transfer cost. This results in the tight electrostatic binding of the ion, and the rigid band doping of the GNM sheet. The pore lattice strictly controls dopant concentration. In the following, we build on this basic physics.
  
In GNM's with nanometer-sized pores, electrostatic dopant binding is expected to decrease due to four effects: 1- A weaker interaction between a dopant and the pore dipoles. 2- The increase in the charge transfer to the sheet, and hence its energy cost. 3- The repulsion between ionized dopants. 4- The introduction of more dopants in large pores can lead to the clustering of dopant atoms. This implies that chemical binding, well known in metals, may occur between the dopant atoms which may affect the doping physics observed with single dopants (Fig. \ref{model}).

In order to understand the implications of these effects on the doping of GNM's with single atomic species, we study the large pore GNM's with multiple dopants for both {\it n}- and {\it p}-doped cases: For {\it n}-type, we consider an oxygen passivated GNM, for two supercells, and two pore sizes. We study the loading of this GNM with multiple lithium and sodium atoms. For {\it p}-type, we consider a hydrogen passivated GNM, also for two supercells, and two pore sizes. The dopants studied here are multiple fluorine and chlorine atoms.
 
We use density functional theory within the generalized gradient approximation to describe the systems of interest as verified in Ref.\cite{Maarouf2013}.  We introduce the nomenclature $n$X@Y--GNM$c$ to denote a GNM structure with a supercell size of $c \times c$ graphene unit cells, with its pore edges passivated by species "Y", and with $n$ atoms of species "X" hosted in the pore. Therefore, our parameter space is constructed by X = \{Li,Na\}, $n = {1,2,..6}$, and $c=9,12$ for the O--GNM systems, and X = \{F,Cl\}, $n = {1,2,..6}$, and  $c=9,12$ for the H-GNM systems. We will cover points in this parameter space that are sufficient to unravel the doping physics. \section{Results and Discussion}

\begin{figure}[H]
\begin{center}
   \includegraphics[width=2.4in]{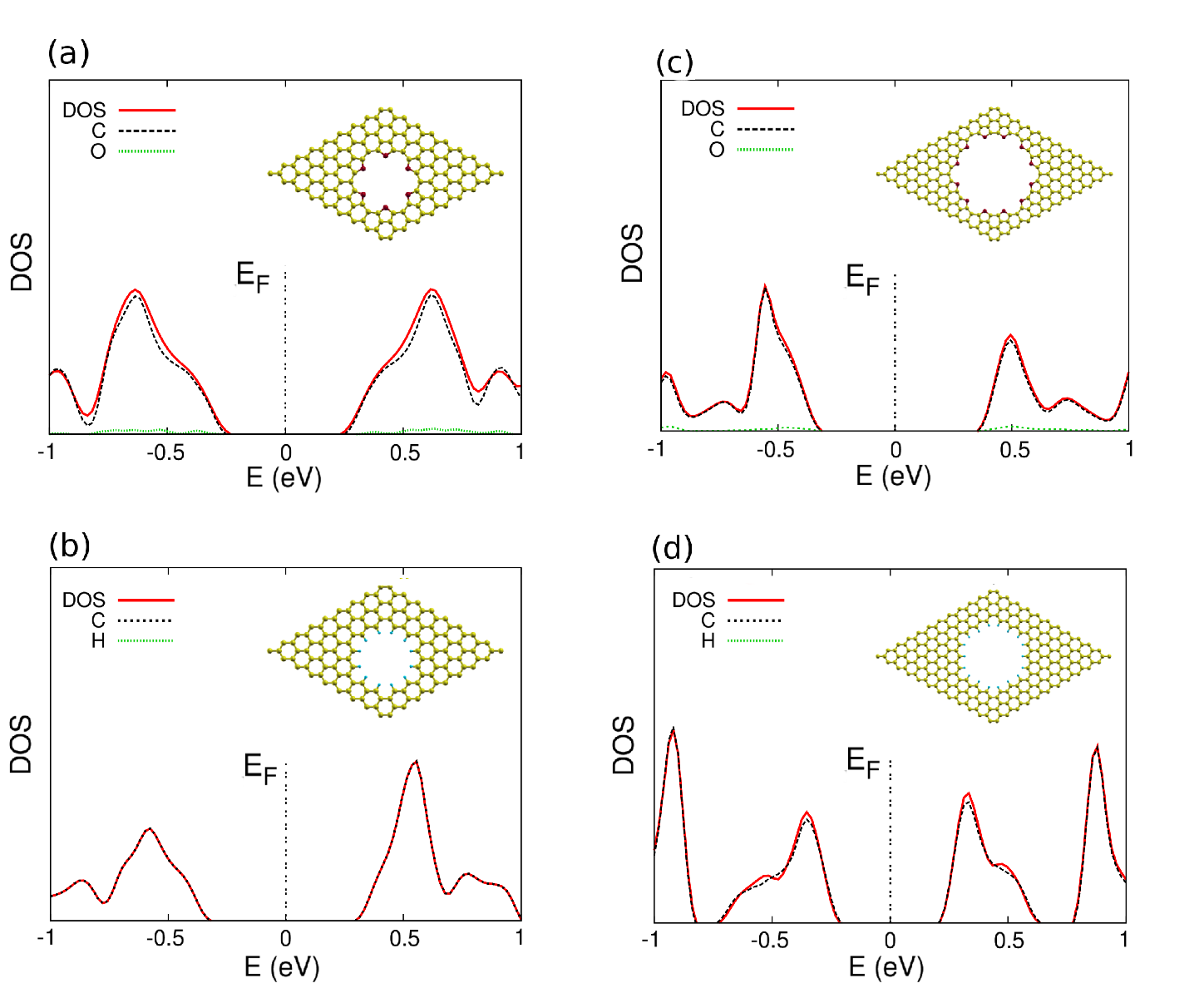}
   \caption{ 
 DOS and PDOS of the pristine (a) O--GNM9, (b) H--GNM9, (c) O--GNM12, and (d) H--GNM12 systems. The insets show the unit cells of the relaxed structures. 
}
   \label{9x9_n_p}
\end{center}
\end{figure}

We begin by presenting the electronic properties of our undoped GNM systems. Figure \ref{9x9_n_p}(a,b) shows an oxygen- and a hydrogen-passivated GNM (insets), with supercell size of $9\times9$ graphene unit cells 
(O--GNM$9$ and H--GNM$9$), as well as their densities of states (DOS). The two GNM's have a pore size of about $0.8$ nm, and are chemically stable as the passivation saturates all bonds. Both GNM's are semiconducting, with a gap of about $0.5$ eV for the oxygen case, and $0.7$ eV for the hydrogen case. Projection of the electronic states on the atomic orbitals (PDOS) shows that the states of the passivating species are located far from the top/bottom of the valence/conduction bands, and are therefore not expected to induce resonant scattering. 

Two GNM's with larger pores are shown in Fig.\ref{9x9_n_p}(c,d) - the oxygen passivated GNM, O--GNM$12$, has a pore diameter of $\sim 1.6$ nm, while the hydrogen-passivated GNM, H--GNM$12$, has a pore diameter of $\sim 1.3$ nm. The DOS of both systems indicate that they are intrinsically semiconducting, with gap sizes of $0.8$ eV and $0.4$ eV for the oxygen-passivated and hydrogen-passivated systems, respectively.

The chemical stability of the studied systems is estimated through a calculation of the zero temperature binding energy ($E_{b}$) of the doped crown GNM system using: 
\begin{equation}
E_{\mathrm{b}} = E_{\mathrm{nX@Y\mbox{-}GNMc}} - E_{\mathrm{Y\mbox{-}GNMc}} - E_{\mathrm{X_n}},
\end{equation}
where $E_{\mathrm{nX@Y\mbox{-}GNMc}}$ is the energy of the $n$X-chelated Y-passivated GNM system with a $c \times c$  supercell size, $E_{\mathrm{Y\mbox{-}GNMc}}$ is the energy of the Y-passivated GNM, and $E_{\mathrm{X_n}}$ is the energy of the most stable molecular cluster of $n$ atoms of species X. Structural relaxation of all systems is started with the dopants at least $4$ \AA\ above the pore. To address the stability of our doped systems, we plot the total energy of a doped system versus the dopants' distance above the pore, measured from their minimum energy position.

\subsection{{\it n}-doping}

In this section, we consider {\it n}-doping cases of O--GNM for $22$ physical embodiments and two dopants; Li and Na, and $2$ pore sizes. This allows us to study cases with multiple dopants in each pore. 

\begin{figure}[H]
\begin{center}
   \includegraphics[width=2.8in]{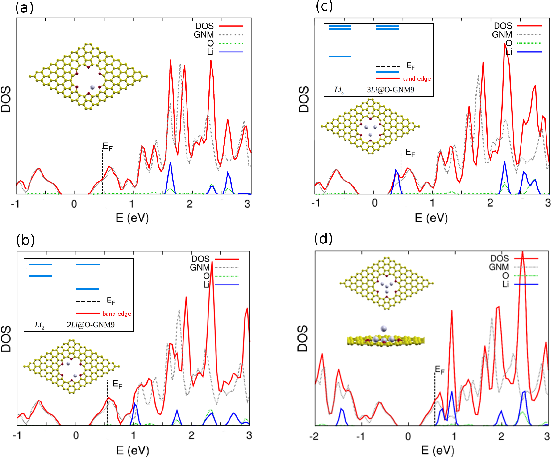}
   \caption{ 
DOS of the (a) Li@O--GNM9, (b) 2Li@O--GNM9, (c) 3Li@O--GNM9, and (d) 4Li@O--GNM9 systems. The bottom insets refer to the relaxed structures. In (b), and (c), the top insets show the Li states within the corresponding O--GNM9 system, as compared to those of the isolated Li clusters. The DOS of the pristine O--GNM9 (GNM) is shown for comparison.}
   \label{Li@O-GNM9}
\end{center}   
\end{figure}

We first consider the cases of O--GNM$9$, with pore diameter $0.8$ nm. A lithium atom brought close to the edge of the pore (inset of Fig.\ref{Li@O-GNM9}a) loses its electron to the graphene skeleton, thereby ionizing and becoming electrostatically trapped in the field of the edge dipoles, forming a Li@O--GNM$9$, with an average Li-O distance of $2.19$ \r{A}, and a binding energy of $1.29$ eV. This is significantly higher than the adsorption energy of Li atop pristine graphene, $\sim 0.4$ eV, as shown by recent studies. \cite{Liu2014} This can be attributed to the electrostatic interaction between the Li and the dipoles formed by the C-O bonds at the edge of the pore. 
 
The DOS and PDOS of the Li@O--GNM$9$ system are shown in Fig. \ref{Li@O-GNM9}a. The Fermi level location indicates that the O--GNM9 is {\it n}-doped. The O--GNM$9$ perturbs the Li state such that it raises its $2s$ state by about $1.1$ eV above the Fermi level. The Li $2${\it s} state is therefore too far up in the spectrum to obstruct low energy electronic transport. By integrating the DOS from the conduction band edge to the Fermi level, we find that one electron has been transferred to the O--GNM$9$. 

We now place another lithium atom close to the pore edge. After structural relaxation, the second atom docks at the pore edge opposite to the first one (inset of Fig. \ref{Li@O-GNM9}b), with a Li-Li separation of $5.28$ \r{A}, and in the same plane of the GNM. The average Li distance to the closest oxygen atoms is $2.10$ \r{A}. The binding energy of the two lithium atoms in the pore is found to be $0.79$ eV. This {\it in-pore} configuration binding energy is about $0.50$ eV higher than that of the system with the second lithium atom {\it atop} the sheet carbon atoms, whether far from the pore, or close to it.

The DOS and PDOS of the 2Li@O--GNM$9$ system are shown in Fig. \ref{Li@O-GNM9}b. The shift in the Fermi energy indicates the increase in the doping level of the $2$Li@O--GNM$9$ over the Li@O-GNM$9$ system. Calculations confirm a charge transfer of $2e$ between the two lithium atoms and the graphene lattice. The DOS shows that the two {\it empty} Li states are now raised above the Fermi level by $0.5$ eV and $1.2$ eV. These two states are the symmetric and antisymmetric linear combinations of the atomic $2${\it s} lithium states. Thus the two Li atoms double the doping level, and maintain the rigid band doping picture, but with a smaller binding energy.

Adding a third lithium atom to the pore introduces a qualitative change. Whereas the states of the single and double Li doping are {\it far} above the conduction band edge of the O--GNM$9$, one of the states of the triple Li doping now falls {\it in} the conduction band of the O--GNM$9$ (Fig. \ref{Li@O-GNM9}c). The O--GNM$9$ perturbs the Li and $2$Li line spectra in such a way so as to raise their atomic/molecular HOMO states, leading to the ionization of lithium atoms in both cases. In the $3$Li case, the perturbation is such that the lowest lithium $2${\it s}-like state lies in the conduction band of the system {\it below} the Fermi level, and therefore two electrons occupy that state, while the third electron is donated to the O--GNM$9$, thus doping it to a level similar to that of the Li@O--GNM$9$ system. The total binding energy in this case is $1.91$ eV. The insets of Figs. \ref{Li@O-GNM9}b and \ref{Li@O-GNM9}c compare the spectra of the two and three Li atoms in the pore to those of the Li$_2$ and Li$_3$ clusters. 


The difference in the binding energies of the Li atoms in the three above cases can be understood as follows: In the Li@O--GNM$9$ case, the cost to ionize the Li atom and to charge the O--GNM$9$ is compensated by the electrostatic energy gain, leading to a binding energy of $1.29$ eV. The $2$Li@O--GNM$9$ system can be thought of starting from a Li$_2$ molecule (dimer) placed in the center of the pore. The pore separates the Li$_2$ to a distance of $5.28$ \r{A}, twice its dimer equilibrium distance of $2.74$ \r{A}, and ionizes the two atoms. The energy cost thus includes the breaking of the Li$_2$ dimer as well as the electrostatic repulsion between the resulting Li$^+$-- Li$^+$ structure. This decreases the binding energy of the 2Li@O--GNM9 to $0.79$ eV. The case of Li$_3$ follows a similar logic, but since the final configuration in the pore has an average Li-Li distance of $3.26$ \r{A}, i.e. nearly the equilibrium distance of the trimer ($2.82$ \r{A}), very little of the electrostatic gain has to be paid towards the trimer deformation. The positive charge of the singly ionized Li$_3^+$ is delocalized over the cluster, causing a stronger electrostatic interaction with the pore oxygen atoms compared to that between the Li$^+$ and its two neighboring oxygen atoms in the Li@O--GNM$9$ case, thus leading to a higher binding energy of $1.91$ eV for the 3Li@O--GNM9 system.

To reach the maximum capacity of the pore, we add more lithium atoms. The fourth atom of the $4$Li@O--GNM$9$ system sits about $2.79$ \r{A} above the plane of the GNM (Fig. \ref{Li@O-GNM9}d). One interesting feature in the $4$Li@O--GNM$9$ system is that the lowest Li$_4$ state is now below the {\it valence} band edge, with {\it two} electrons transferred from the cluster to the O--GNM. The binding energy of the $4$Li@O--GNM$9$ is $1.59$ eV. At higher doping ($5$Li@O--GNM$9$, and $6$Li@O--GNM$9$), a Li state persists in the conduction band of the system (supplementary data, Fig. S1).
 
\begin{figure}
\begin{center} 
   \includegraphics[width=3.0in]{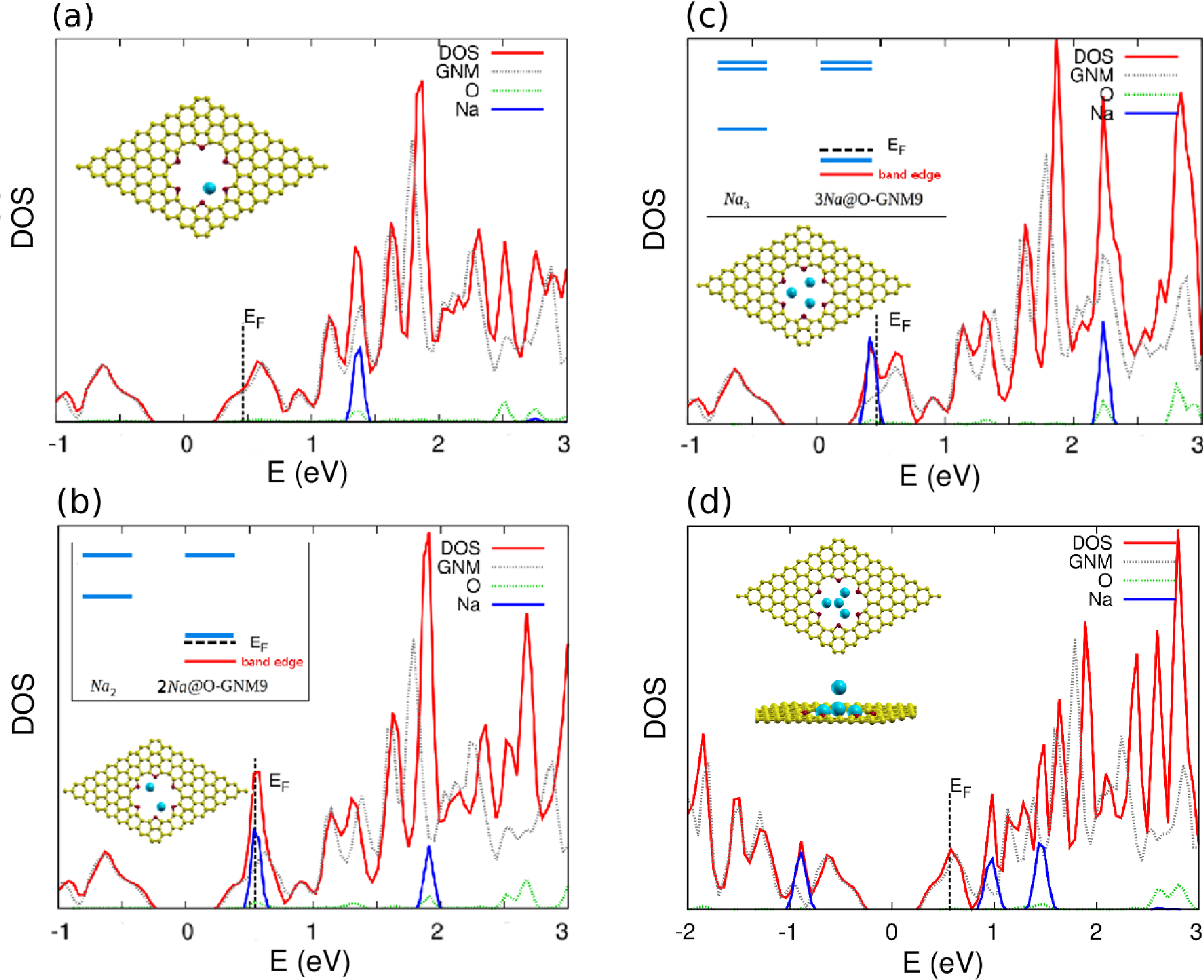}
   \caption{ 
  DOS of the (a) Na@O--GNM9, (b) 2Na@O--GNM9, (c) 3Na@O--GNM9, and (d) 4Na@O--GNM9 systems. The insets refer to the relaxed structures. In (b), and (c), the top insets show the Na states within the corresponding O--GNM9 system, as compared to those of the isolated Na clusters. The DOS of the pristine O--GNM9 (GNM) is shown for comparison.}
\label{4567LiNa}  
\end{center}   
\end{figure}  
  
\begin{figure}
\begin{center} 
   \includegraphics[width=3.0in]{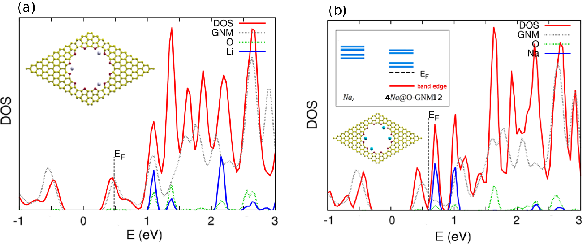}
   \caption{ 
  DOS of the (a) 3Li@O--GNM12, and (b) 4Na@O--GNM12 systems. The bottom insets refer to the relaxed structures. In (b), the top inset shows the Na states within the corresponding O--GNM12 system, as compared to those of the isolated Na$_4$ cluster. The DOS of the pristine O--GNM12 (GNM) is shown for comparison.
}  
 \label{Li@O-GNM12}  
\end{center}   
\end{figure}

              
We now turn to {\it n}-doping of the larger pore, where we consider six systems: Li@O--GNM$12$, $2$Li@O--GNM$12$, $2$Na@O--GNM$12$, $3$Li@O--GNM$12$, $4$Na@O--GNM$12$, and $6$Na@O--GNM$12$. For 1 and 2 dopant atoms, the behavior is qualitatively similar to the small pore case (supplementary data, Fig. S4). For 3 and 4 dopants, the situation is different. Figure \ref{Li@O-GNM12}a shows the structure and DOS of the $3$Li@O--GNM$12$ system, indicating that all three Li atoms are now ionized, with 3 electrons transferred to the GNM. The system has a binding energy of $1.39$ eV. All three Li-like states are now above the Fermi energy.

The higher load case of 4Na@O--GNM$12$ is shown in Fig. \ref{Li@O-GNM12}b. All Na atoms are ionized, contrary to the small pore 4Na@O--GNM$9$ system, with 4 electrons now doping the GNM. The O--GNM12 pulls down the lowest two states of the Na$_4$ cluster, positioning them just above the Fermi level of the system, signaling that higher loads will result in pushing Na states down into the conduction band of the GNM. Indeed, with 6 Na atoms, the 6Na@O--GNM$12$ system has its lowest Na state {\it below} the Fermi energy, now with only 4 electrons doping the GNM (supplementary data, Fig. S4).            
                       
\begin{figure}[H]
\begin{center} 
    \includegraphics[width=3.0in]{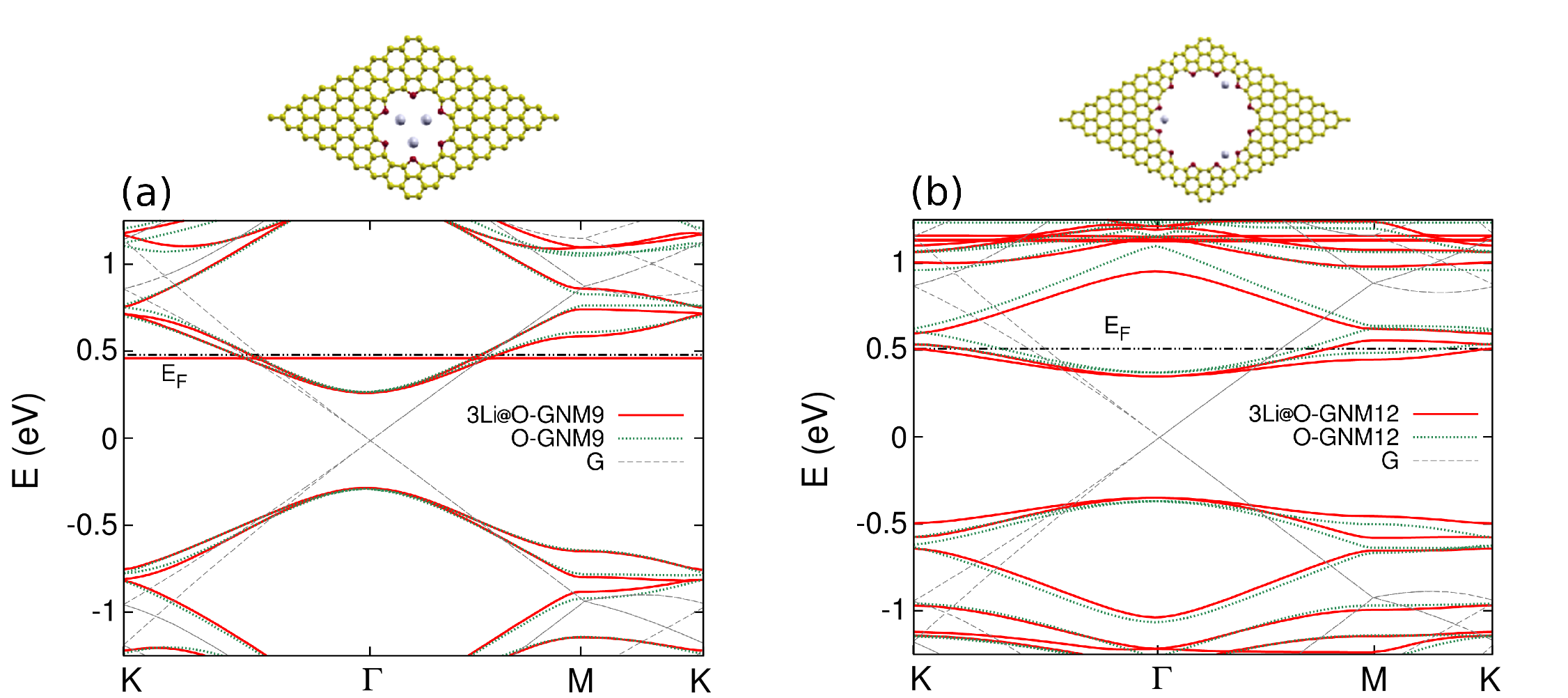}
   \caption{ 
  Electronic band structures of (a) 3Li@O--GNM9, and (b) 3Li@O--GNM12 systems. A Li state lies just below the Fermi energy in the 3Li@O--GNM9 case. The band structure of graphene, pristine O--GNM9, and O--GNM12 are shown for comparison. 
}
 \label{n_bandstruct}  
\end{center}   
\end{figure}

To confirm the rigid band nature of the doping, we show the band structures of two {\it n}-doped systems; 3Li@O--GNM9 (Fig.\ref{n_bandstruct}a), and 3Li@O--GNM12 (Fig.\ref{n_bandstruct}b). Compared to the undoped O--GNM's, there is no significant change in the band curvatures in the Fermi level/gap region. The pristine graphene is also shown for comparison. The bands are linear in the vicinity of the Fermi energy, with a group velocity of about half/quarter that of graphene for the O--GNM9/O--GNM12, respectively.  

Now that we have understood the effect of the dopants on the electronic structure of all {\it n}-doped systems, we compute the free energy differences for various reactions forming them. This is needed as the entropic cost for forming these systems may not be negligible. We will assume that our systems are formed according to:
\begin{equation}
2~(n-1)\text{X@O--GNM} ~ + ~ \text{X}_2 ~ \rightarrow ~  2~n\text{X@O--GNM},
\end{equation}
where $n$ indicates the number of dopants, and X=Li,Na. We calculate the free energy difference
\begin{equation}
\Delta A = A_{\text {reactants}} - A_{\text {products}}
\end{equation}
of the formation of all systems. With this definition, a positive free energy difference indicates that it is favorable to add one more dopant.

\begin{figure}[H]
\begin{center} 
\begin{tabular}{c}
  \includegraphics[width=2.9in]{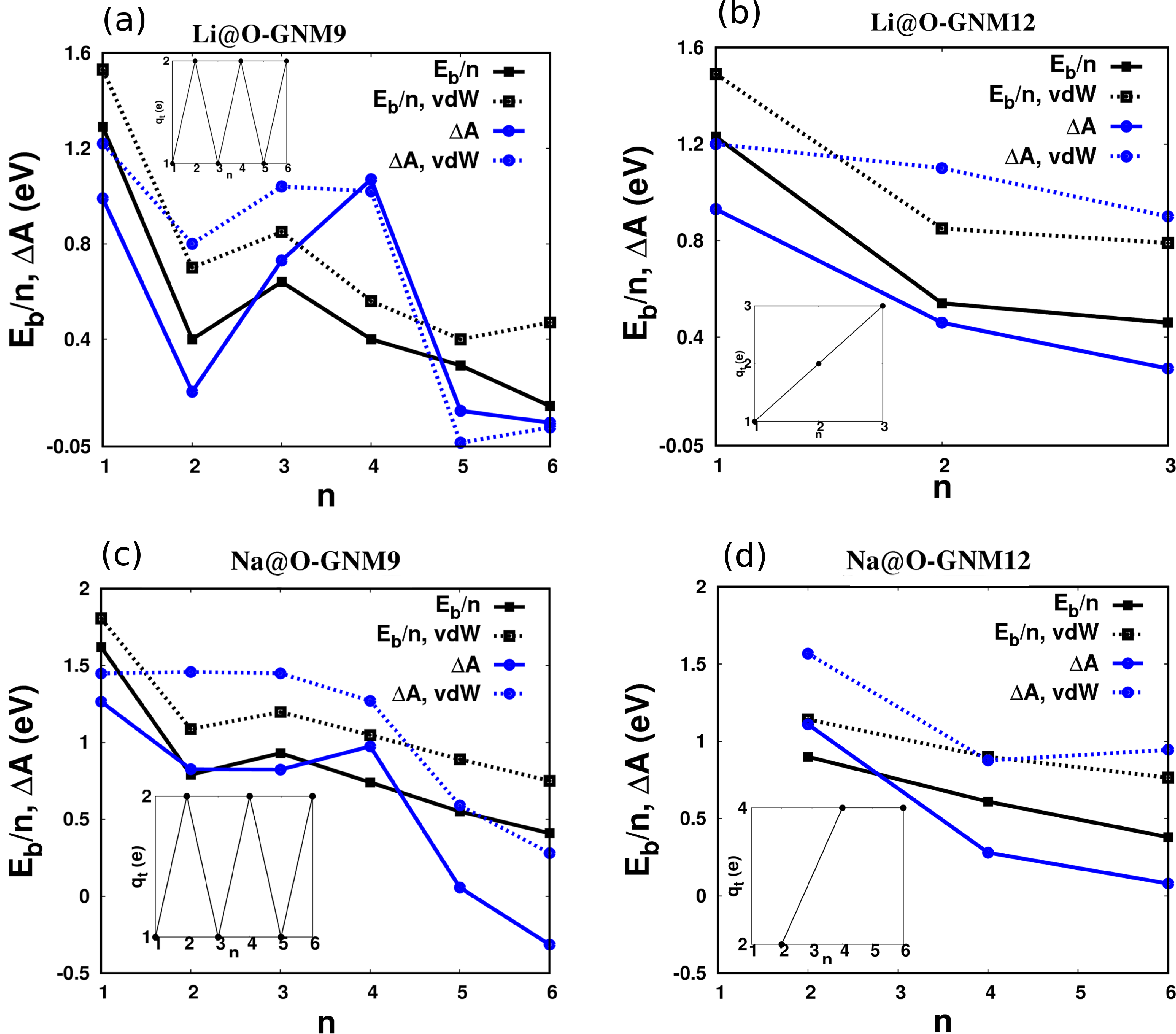}
\end{tabular}
\caption{Binding energies per dopant ($E_{b}/n$, eV), and free energy difference ($\Delta A$, eV) (with and without vdW) of various {\it n}-doping systems. The insets show the charge transferred ($q_t$, electrons).
} 
\label{ndopingtable1}
\end{center}   
\end{figure}

The pore edges of GNMs are essentially composed of radial dipoles, and since the dopants extend spatially over the pore region, it is instructive to study the effect of the vdW corrections to the calculated energies. Figure \ref{ndopingtable1} shows the free energy difference  $\Delta A$, the binding energy per dopant $E_{b}/n$ (with and without vdW), and the charge transfer for all {\it n}-doped O--GNM structures. Although the vdW corrections are small for light doping, they can be significant for high load. For the small pore systems, the corrections range between $\sim 0.2$ and $1$ eV, while they are higher for the large-pore O--GNM12 systems ($\sim 0.5$ to $2$ eV). The higher effect in the large pore case can be attributed to the increased spatial distribution of the charges in the pore region compared to the smaller-pore O--GNM9 systems.

In general, the Na systems have the advantage of having a higher binding energy per dopant. On the other hand, the formation of 4Li@O--GNM9 is favored over higher doping within the Li systems, which is not the case with Na. Since this structure has no Li states in the conduction band, doping with Li could be an advantage for a GNM-based transistor device. 
   
\subsection{{\it p}-doping}
We now examine the {\it p}-doping of large pore GNM systems. We consider 6 doped H--GNM$9$ systems (pore size 0.8 nm), and 4 doped H--GNM$12$ systems (pore size 1.3 nm), using F and Cl. The reported binding energies are calculated from the energies of the most stable form of the dopants (F$_2$ and Cl$_2$ molecules). 

\begin{figure} 
\begin{center}
   \includegraphics[width=3.0in]{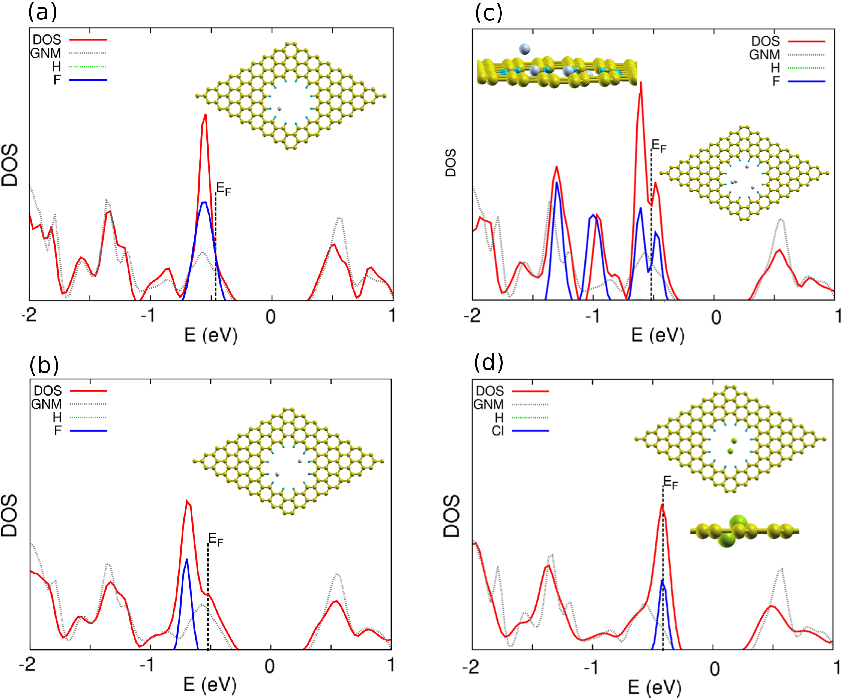}
   \caption{ 
  DOS of the (a) F@H--GNM9, (b) 2F@H--GNM9,(c) 4F@H--GNM9, and (d) 2Cl@H--GNM9 systems. The inset figures refer to the relaxed structures. In (c) the inset shows a side view of the 4F@H--GNM9 system, indicating the off-plane structure of the 4F dopants. The DOS of the pristine H--GNM9 (GNM) is shown for comparison.
}
   \label{F@H-GNM9}
\end{center}   
\end{figure}

A fluorine atom brought close to the edge of the H--GNM$9$ pore will have its spectrum perturbed by the pore dipoles, resulting in the LUMO of the F falling below the valence band edge of the GNM. This will cause an electron to spill from the GNM skeleton to occupy the F LUMO, thus ionizing it, and {\it p}-doping the GNM. The electrostatic attraction with the nearby hydrogen atoms creates an ultra stable environment for the F anion. Figure \ref{F@H-GNM9}a shows the structure and DOS of the F@H--GNM9. The average F--H distance with the nearest two hydrogen atoms is $1.70$ \r{A}. The binding energy of F@H--GNM$9$ is found to be $1.07$ eV. The H--GNM causes the F $2${\it p$_z$} state to be pinned about $0.3$ eV below the valence band edge of the H--GNM, and $0.1$ eV below the Fermi level.

The structure and DOS of the $2$F@H--GNM$9$ system is shown in Fig. \ref{F@H-GNM9}b. The GNM is {\it p}-doped, with the Fermi energy located $\sim 0.3$ eV below the valence band edge. The binding energy of the system is $1.72$ eV. The two F atoms are now ionized, with their $2${\it p$_z$} states full and located $\sim 0.2$ eV {\it below} the Fermi level. Doping with 3 fluorine atoms leads to a qualitatively similar scenario (supplementary data, Fig S5a).

To reach the maximum pore capacity of the H--GNM$9$, we add a fourth fluorine atom to form $4$F@H--GNM$9$. The system reaches its minimum energy by having two of its fluorine atoms form an {\it elongated} fluorine molecule, F$_2^*$, which protrudes out of the pore plane (Fig. \ref{F@H-GNM9}c), with a bond length of $1.68$ \r{A} (a F$_2$ molecule has a bond length of $1.43$ \r{A}). The F$_2^*$ is $2.21$ \r{A} from the closest pore-edge hydrogen. The DOS of the $4$F@H--GNM$9$ system shows that there is one fluorine state just {\it above} the Fermi energy. Inspection of this state indicates that it is a superposition of the $2${\it p$_z$} states of the two fluorine atoms forming the elongated molecule. These two fluorine atoms are {\it not} ionized. This is confirmed by the L\"owdin charge analysis, which also shows that the other two fluorine atoms are each singly ionized. The binding energy of the $4$F@H--GNM$9$ system is $2.02$ eV, which is only $\sim$ $0.10$ eV higher than that of the $3$F@H--GNM$9$ system, suggesting that $3$F@H--GNM$9$ is the maximally-doped and most stable fluorine-doped H-GNM9 structure, and hence best suited for a {\it p}-doped H--GNM$9$-based device. An energy scan of $3$F@H--GNM$9$ system where the fluorine dopants are moved above the pore indicates that no other minima exist in the pore vicinity (supplementary data, Fig S6). 

We now switch to the other {\it p}-dopant, chlorine. Doping with one Cl atom gives a scenario similar to doping with one F atom (supplementary data, Fig. S5b). Upon adding another Cl atom, the system favors the formation of an {\it elongated} Cl$_2$ molecule, Cl$_2^*$, which sticks out of the plane of the GNM (Fig. \ref{F@H-GNM9}d). The Cl$_{2}^*$ molecule has a Cl--Cl distance of $2.28$ \r{A}, larger than the Cl$_2$ molecular distance of $2.00$ \r{A}. The system has a small binding energy of $0.19$ eV. The DOS of the system shows that the GNM is slightly doped and that the Cl 3{\it p$_z$} states are only partially full. The L\"owdin charge analysis confirms this picture showing that the Cl$_2$* structure carries a small negative charge of about $0.2e$. Adding more Cl dopants does not lead to stable structures.

\begin{figure}[H]
\begin{center} 
         \includegraphics[width=3.0in]{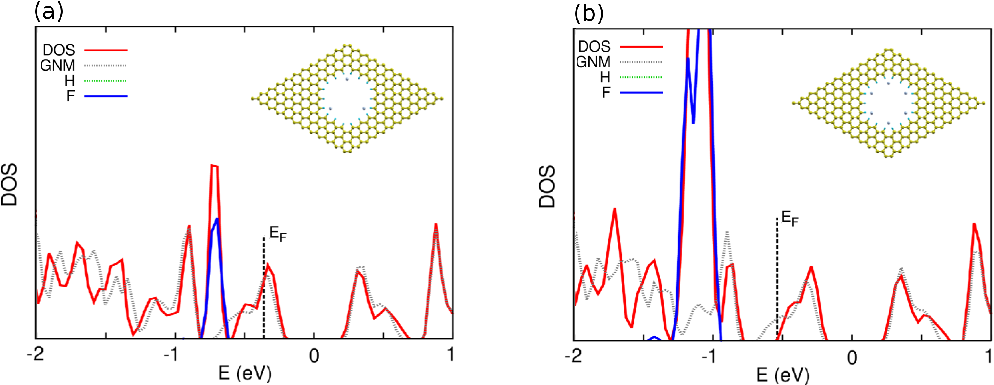} 
   \caption{ 
   DOS of the (a) 3F@H--GNM12, and (b) 6F@H--GNM12 systems. The inset figures show to the relaxed structures. The DOS of the pristine H--GNM12 (GNM) is shown for comparison.
}
   \label{F@H-GNM12}
\end{center}   
\end{figure}

Next, we describe our {\it p}-doping results for larger pores. We form a hydrogen doped GNM with a pore size of about $1.3$ nm, H--GNM$12$. Since the pore is large, and three fluorine atoms can stably dope the smaller pore system, we begin with $3$F@H--GNM$12$ (Fig. \ref{F@H-GNM12}a) The average distance between the fluorine and the closest two hydrogen atoms in the structurally relaxed system is $1.67$ \r{A}. The average F--F distance is $8.62$ \r{A}. The DOS of the system shows that the GNM$12$ is {\it p}-doped, with the Fermi energy placed $\sim 0.1$ eV below the valence band edge. The fluorine $2${\it p$_z$} states are all full, and are located at least $0.3$ eV below the Fermi level. The integrated DOS shows that the GNM lost three electrons. The binding energy of the $3$F@H--GNM$12$ is $2.72$ eV.

Adding more fluorine atoms leads to similar physics. Doping with 4 and 6 fluorine atoms removes 4 and 6 electrons from the GNM, respectively, confirming the full ionization of the fluorine atoms in both systems. We only show the $6$F@H--GNM$12$ system here (Fig. \ref{F@H-GNM12}b. The details of the $4$F@H--GNM$12$ case are presented in the supplementary data, Fig. S7a). The highest filled fluorine $2${\it p$_z$} state is $\sim 0.5$ eV away from the Fermi level, which is pinned $0.3$ eV below the valence band edge. The binding energy of the $6$F@H--GNM$12$ is $2.76$ eV. Finally, Cl doping of the larger pores gives slightly more stable structures compared to small pore systems, albeit with small binding energies (Fig. S7b). This further confirms that fluorine should be the dopant of choice for H--GNM systems.

\begin{figure}[H]
\begin{center} 
\includegraphics[width=3.0in]{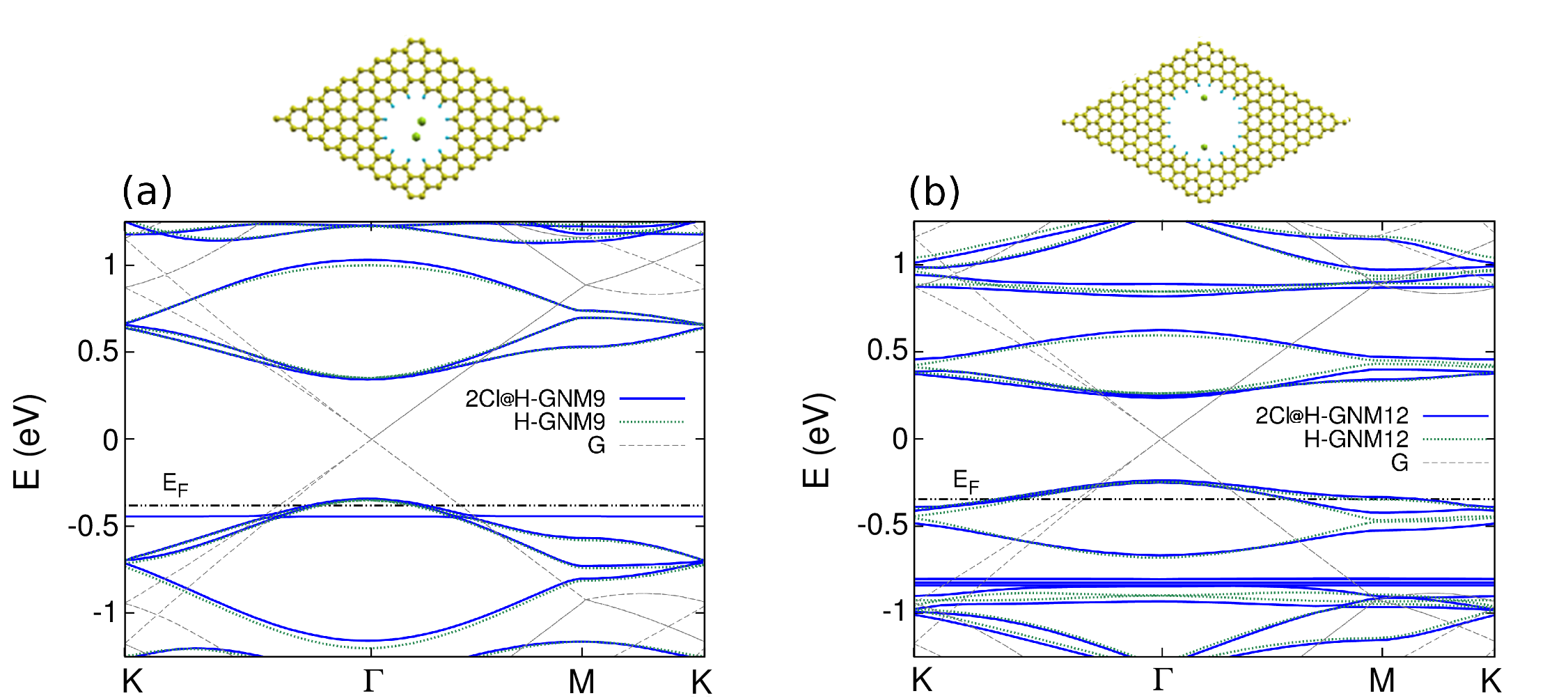}
\caption{Electronic band structures of (a) 2Cl@H--GNM9, and (b) 2Cl@H--GNM12. The band structure of graphene, pristine H--GNM9, and H--GNM12 are shown for comparison.}
\label{p_bandstruct}  
\end{center}   
\end{figure}

Chelation {\it p}-doping of GNM's occurs in a rigid band manner, as is demonstrated by the band structure of the doped systems. In Fig. \ref{p_bandstruct}, we show the band structures of the 2Cl@H--GNM9 and 2Cl@H--GNM12 systems. Band curvatures are very similar to those of the pristine GNM systems. In addition, they are linear in the neighborhood of the Fermi level, with electronic group velocity half/quarter that of graphene for the H--GNM9/H--GNM12. 

\begin{figure}[H]
\begin{center} 
   \includegraphics[width=3.3in]{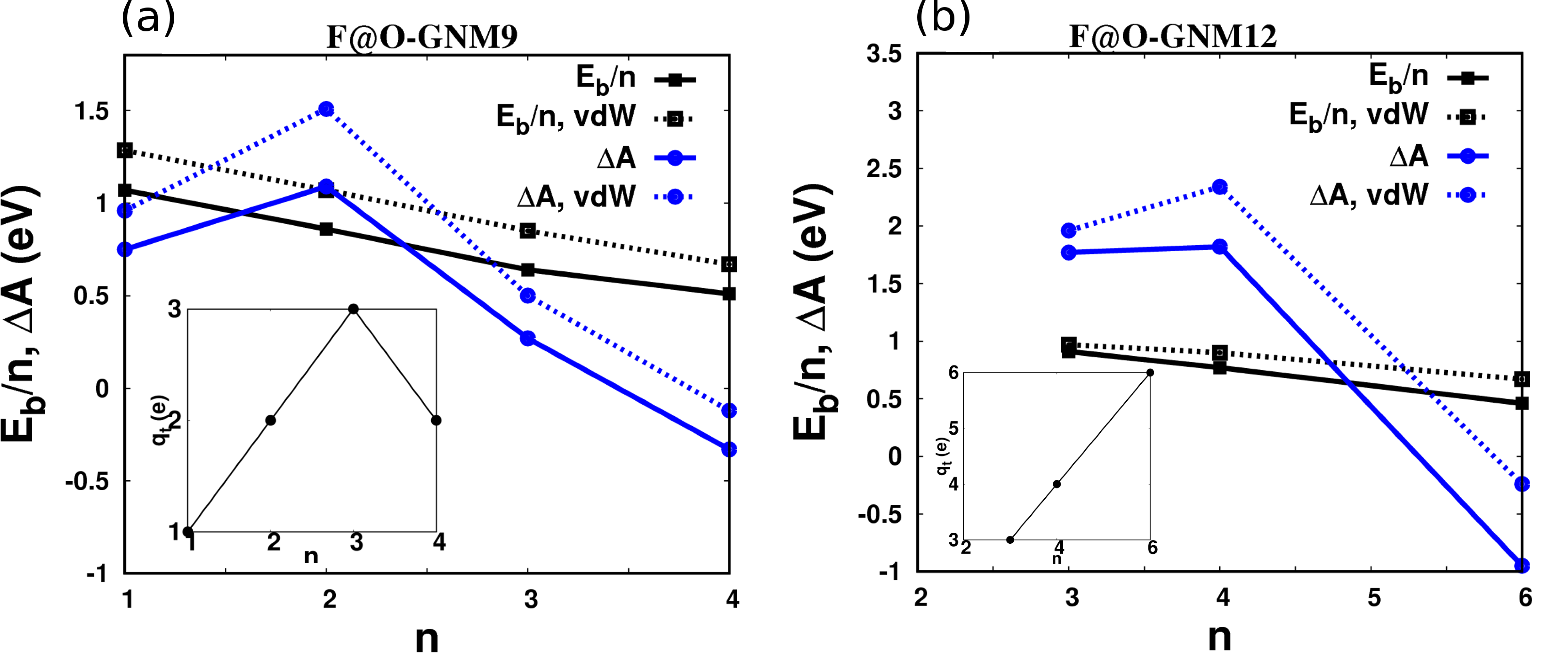}
   \caption{Binding energies per dopant ($E_{b}/n$, eV), free energy difference ($\Delta A$, eV) (with and without vdW) of various F {\it p}-doping systems. The insets show the charge transferred ($q_t$, electrons)
}
\label{pdopingtable1}
\end{center}   
\end{figure}

As in the {\it n}-doping case, we take into account the entropic cost of forming our {\it p}-doped systems. We calculate the free energy for the reaction forming the structure $(n+1)$F@H--GNM from the lesser doped structure $(n-1)$F@H--GNM:
\begin{equation}
(n-1)\text{F@H--GNM} ~ + ~ \text{F}_2 \rightarrow ~  (n+1) \text{F@H--GNM},
\end{equation}
where $n$ indicates the number of F dopants (chlorine is excluded due to its weak binding to the pores), with and without vdW. As we see in Fig. \ref{pdopingtable1}, the 3F@H--GNM9 system is the most likely to occur for {\it p}-doping the H--GNM9 with fluorine. For the larger pore systems, 4F@H--GNM12 is the most probable system. As in the case of the {\it n}-dopants, the vdW corrections appear to be significant, and they range between $\sim 0.2$ and $0.4$ eV for small pores systems, and $\sim 0.2$ and $1.5$ eV for large pore ones. 

Based on our results of the two studied pore sizes, we can infer how the doping occurs with larger ones. For the {\it n}-doping case, dopants will first bind to the pore edges. As the load is increased, clusters will form in the pore, giving rise to similar doping physics discussed in this work. For maximum loading, the number of dopants grows cubically with the cluster/pore radius, whereas the number of binding sites increases only linearly. As a result, the binding energy {\it per} dopant will decrease as the cluster size increases. On the other hand, the {\it p}-dopant, fluorine, will continue to bind to the pore edge with increasing pore radius.  
  
Fabrication of GNM pores is bound to produce pores with some level of disorder. As a result, a GNM will have a distribution of pores sizes and pore lattice constants, changing the symmetry requirement for the electronic gap formation. The effect of this is two-fold. First, the gap will partially close, and second, the dopant load will vary for different pores. The latter effect will have negligible impact on potential applications of the GNM's, as the resulting structures are still stably doped, whereas the former can seriously prevent the development of GNM-based transistors with technologically acceptable ON/OFF ratios.   
 
Chelation-doped crown GNM's may be considered for many other applications, such as transparent electrodes. On the other hand, their use in microelectronic applications requires controlled doping of the sheets. Thus the pores, post fabrication, must be filled with the desired number of dopants, no more, no fewer. To achieve this result, it is convenient to select the binding energy dopant level with the lowest free energy such that sufficient thermal annealing, for instance, will allow the GNM/dopant system to relax to the correct number of dopants per pore. This approach avoids difficult stoichiometric control and other problems associated with achieving a doping state that is not the most thermodynamically stable. There are two caveats. First, choosing the right dopant whose lowest free energy state is a good fit for microelectronics. Second, the free energy of the next most stable dopant should be well separated from the lowest stable (large free energy gap) such that the thermal anneal will be effective. The results indicate that to achieve optimal control in transistor applications for large pores, dopants must be designed to fit in the pore of interest, allowing for carefully controlled device engineering.
  
\section{Conclusion}

Graphene nanomeshes (GNM's) with easily-fabricated pores ($\sim$ 1 nm) have high transparency, fractional band gaps, and electronic mobilities that are attractive for many applications. Their utilization requires a mechanism for their stable doping. We use first-principles calculations to show that it is possible to achieve stable controlled doping of large-pore GNM's using multiple ion chelation. Selective passivation of the pore edges allows for the hosting of multiple dopants, which are ionized and electrostatically anchored in the pores, thereby doping the GNM.  

Stable {\it n}-doping of a $0.8$ nm pore GNM using multiple dopants is possible using Li$_n$, and Na$_n$ clusters, $n=1..6$, with a binding energy that is higher than $50k_BT$ at room temperature. For larger pores ($\sim 1.3$ nm), more dopants can be hosted, with a binding energy of the same order. For the case of {\it p}-doping, stable rigid band doping occurs with up to 4 F atoms in the first hydrogen-passivated pore, with the three F case being the most energetically favorable. Chlorine doping does not lead to stable doped structures. The larger pore systems also favor Fluorine as a dopant, where 6 F atoms can be stably hosted, while chlorine doped systems remain unstable. We also find that doping occurs in a rigid band way for both doping flavors, and states at the Fermi level of the system are the graphene carbon {\it p}$_z$ states. Our results also show that van der Waals corrections contribute significantly to the binding energies, ranging from $\sim 30-40 \%$, to $\sim 90 \%$ for some {\it n}-doping cases with few-atom clusters.

Our results satisfy the need to study relatively large systems with bigger pores, and more ions per pore. This renders GNM's closer to an experimentally realizable device in terms of length scale, where they may be considered for use as transparent electrode and field effect transistors. However, it seems that large complex molecular dopants, designed for the desired pore sizes are indicated for microelectronics applications such that the charge transferred to the GNM can be easily controlled.

\section*{Acknowledgment} 
\vspace*{-0.3cm}
A. Maarouf would like to acknowledge the support of the supercomputing facility at the Bibliotheca Alexandrina, Alexandria, Egypt, and in part, the resources of the Supercomputing Laboratory at King Abdullah University of Science \& Technology (KAUST) in Thuwal, Saudi Arabia, and the resources and technical services provided by the Scientific and High Performance Computing Center at Imam Abdulrahman Bin Faisal University, Dammam, Saudi Arabia. A. Maarouf and M.\ Fadlallah\ would also like to thank Ulrich Eckern and Udo Schwingenschl\"ogl for fruitful discussions.

\section*{Appendix A. Supplementary data}
Supplementary data associated with this article can be found, in the online version, at ..

\section*{References}
\bibliographystyle{elsarticle-num}
\bibliography{2revised_dopingporousgraphene}


\end{document}